\documentclass[12pt]{article}
\usepackage{amsmath}
\usepackage{amsfonts}
\usepackage{amssymb}
\usepackage{latexsym}
\usepackage{graphicx}
\usepackage[english]{babel}
\input epsf.sty

\begin{document}

\title{Typicality, Freak Observers and the Anthropic Principle of Existence}

\vspace{3mm}
\author{{Miao Li$^{1,2}$\footnote{mli@itp.ac.cn}, Yi Wang$^{2,1}$\footnote{wangyi@itp.ac.cn}}\\
{\small $^{1}$ The Interdisciplinary Center for Theoretical Study,}\\
{\small University of Science and Technology of China (USTC), Hefei,
Anhui 230027, P.R.China}\\ {\small $^{2}$ Institute of Theoretical
Physics, Academia Sinica, Beijing 100080, P.R.China}}
\date{}
\maketitle

\begin{abstract}
We propose an alternative anthropic probability for calculating the
probabilities in eternal inflation. This anthropic probability
follows naturally from the weak anthropic principle, and does not
suffer the freak observer or the typicality problems. The
problem that our observed cosmological constant is not at the peak
of the usual anthropic probability distribution is also solved using
this proposal.

\end{abstract}

\newpage

\section{Introduction}

Recent work in string theory indicates that there may be a vast
landscape of meta-stable vacua in the string theory
\cite{landscape}. Some constants of nature are used to label these
vacua thus can not be uniquely determined from the first principle.
This string theory landscape, just like any multitude of vacua, can
be realized in cosmology through the eternal inflation scenario
\cite{steinhardt-nuffield,vilenkin-eternal,linde-eternal}. If this
is to be the case, we can not expect to have a deterministic
explanation of nature, we can only calculate probability for each
vacuum. The measure or the probability distribution  in eternal
inflation have become one of the central problems in cosmology
\cite{vilenkin-measure, Bousso:2006ev, Gibbons:2006pa, Linde:2006nw,
Podolsky:2007vg, Li:2007rp, Cai:2007et}. Whether the eternal
inflation scenario can make any useful prediction, even less precise
in the eyes of traditional science, is a matter of debate.

Unfortunately, if we do not use extra selectional effects in
calculation of probability, vacua with large probability never look
like the vacuum we live in. To take a selectional effect into
account one usually starts with Bayes' formula
\begin{equation}
  P({\rm theory}~x|{\rm selection})=\frac{P({\rm selection}|{\rm theory}~x) P({\rm theory}~x)}{\sum_y P({\rm selection}
  |{\rm theory}~y) P({\rm
  theory}~y)}~,
\end{equation}
where the word selection stands for a set of selectional conditions.

The anthropic principle \cite{Weinberg:1988cp} is used as a kind of
selectional conditions in this calculation of probability, although it betrays the usual
concept of scientific selectional principles-such as the lowest energy
principle. In addition to
the anthropic principle, there are other possible selectional effects such as those discussed in
\cite{Smo04}. There is also the possibility that no such selectional
effect really works and some constants of nature are determined by
pure chance.

One of the authors of the present note (ML) is fully aware of the fact that
in the history of science many false leads appeared because scientists are
blinded by their limited experimental knowledge as well as limited theoretical understanding
at the time, chasing after aether is one of such examples. To the 19th century theoretical
physicists, aether seemed to be inevitable. Anthropic selectional effects, as applied to
some of nature constants, may be another false lead. Nevertheless, so long as it appears
inevitable at the present time, it is a legitimate scientific topic to study.

There have been several versions of the anthropic principle
\cite{Weinberg:1988cp}, for example, the strong, the weak, and the weakest
anthropic principle. The weakest anthropic principle states that
``our observations about the universe should be in accord with the fact that we are observing it''.
In this weakest version, the selectional effect is no
more than a set of experimental data. It does not try to explain the
coincidence problems for the existence of observers
\cite{Weinberg:2005fh}, unless the solution of coincidence is
contained in the theory itself. Recently, Hartle an Srednicki
\cite{Hartle:2007zv} discussed the typicality problem and method
of calculation using the weakest anthropic principle. While other
discussions on this issue can be found in \cite{Page:2007bt}.

In this paper we mainly discuss the weak anthropic principle, all it
says is  ``what we can observe is restricted by the conditions
needed for the presence of observers." The weak anthropic principle
is the weakest one which can explain the coincidences for the
existence of observers, so it provides a necessary and modest
starting point for our discussion. The recipe we propose can help to
avoid several puzzles concerning the current study in this
direction, the main result of our proposal is that among those
theories (or vacua) which can accommodate a large number of
galaxies, the preferred one (ones) is largely determined by the a
priori probability.

When we want to calculate probabilities using the anthropic
principle, we need to transcribe this principle into formulas.
Several kinds of anthropic probability are proposed, such as
the volume weighted, galaxy weighted \cite{vilenkin-measure,
Linde:2006nw} and entropy weighted \cite{Bousso:2007kq}
probabilities. A common attribute of these probabilities is that
they are proportional to the number of the observers. So in the
following discussion, we call them number weighted probability
uniformly. In the number weighted method, the probability is
expressed as \cite{vilenkin-measure}

\begin{equation}
  P_{\rm obs}(X)\sim P(X)n_{\rm obs}(X),
\end{equation}
where $P(X)$ is the volume fraction of interest, for example, the
volume for thermalized regions, with the given values of the
constants $X$. And $n_{\rm obs}(X)$ is the number of observers (or
the amount of entropy production) in such regions per unit volume.
In such a number weighted prescription, the freak observer problem
\cite{bb} often arises, and the typicality of our human observer is
debated \cite{typicality1, Hartle:2007zv, Page:2007bt}.

In this paper, we propose an alternative formula to calculate the
anthropic probability. In Section 2, we give our recipe from
anthropic reasoning. It is shown that this probability follows
directly from the anthropic principle, and does not suffer the
typicality problem or the freak observer problem. In Section 3, we
discuss the derivation-from-peak problem of the observed
cosmological constant as an application of this probability. We also
use a simple model to illustrate the use of this probability
combined with the a priori probability. We conclude in Section 4.

\section{The anthropic principle of existence}

In this section, we propose an alternative anthropic probability.
The anthropic principle claims that ``what we can expect to
observe must be restricted by the conditions necessary for our
presence as observers''\cite{anthropicexpression}. So it is natural
that the probability should not be proportional to the number of
observers, rather, it is just the probability for the existence of observers.
For example, if the probability for observers to appear per
galaxy (or space volume in a given unit) is $p$, and there are $n$
galaxies (or space volumes) in our pocket universe, then instead of
the number weighted probability $np$, the anthropic probability
should be
\begin{equation}
  P(p,n)=1-(1-p)^n~.\label{ap}
\end{equation}
It is clear that when $np\ll 1$, $P(p,n)$ is approximated by $np$.
If observers are not rare in galaxies, this probability becomes
almost one, and a further increase of number of galaxies does not
significantly change the probability. This property is very
different from that of the number weighted probability. To get the
full probability using  Bayes' formula, one also needs to multiply a
priori probability for a given theory.

Besides reasoning from the definition of anthropic principle, there
are also several other reasons that the existence-or-not weight is
more natural then the weight given by the number of the observer.

First, the main support for the anthropic principle is the
coincidence that we just live in the universe suitable for us to
live in. The existence of the observer is needed in this argument,
while the number of the observers does not figure in.

Second, there is ``experimental" evidence that we are not selected
by the number of observers. For example, anthropic reasoning can be
used to explain why we live on the earth where there
is liquid water. But the earth does not seem to be the largest planet
in the universe suitable for human to live, where more human beings
can be developed. On the other hand, we do not seem to be the
smallest possible intelligent beings in size suitable to live on the earth
\cite{PressPage}, which consumes less resources, so can develop a
larger population. In conclusion, the anthropic selectional effect does
not become stronger just because there can be more observers.

Third, there are paradoxes in the number weighted probability, which
do not arise in our probability.

One paradox is the freak observer problem (also known as the
Boltzmann brain problem). If the universe is asymptotically de
Sitter, there should be an infinite number of observers developing from the
thermal fluctuations. This is more than the finite number of
observers like us. So the question is, why we are human observers,
not freak observers.

Calculation using the number weighted measure suffers this
problem in two or three ways. First, it is often assumed in such
calculation that we are typical observers, because the probability
for a species of observer to do observation is assumed to be
proportional to the number of this species of observers. But if freak
observers are infinite, we can not be typical. Second, if there are
both a finite number of humans and a finite number of freak observers (or without freak
observers) in our universe, then the anthropic probability for our
universe should be infinitely small compared with some other
universe with infinite number of freak observers, which can be
self-consistently realized by a pure de Sitter phase, or something
like that.

The anthropic probability we propose suffers none of these two
problems. The first problem is bypassed because we never assume whether we
are typical or not in our reasoning. The selectional effect is always
a selection of vacua, but not a selection of observers. The second
problem is avoided because if there are a large number of human
observers in the universe, then the corresponding anthropic
probability is nearly one. So it is not much smaller than the
probability for a universe filled by a infinite number of freak
observers.

There is potentially a third kind of the freak observer problem that
if the freak observers really exist and require less meticulous arrangements
(coincidences) than
the human observer, then the anthropic explanation for the
coincidences is pointless. The anthropic probability we propose is
not completely free of this kind of freak observer problem. But it
still fares better than the number weighted anthropic probability
in this regard. On the other hand, it is not clear whether or not the
assumption that the freak observer requires less coincidences than the
human observer is correct. So we do not discuss this problem in
detail in this paper.

Another paradox is the prediction of our fate. As the population is
now growing on the earth, why do we live at the present time, while
do not live in the future? Does it predict that the population will
stop growing in the near future? It seems that such prediction is
absurd, and the above questions should be answered by other
branches of science such as sociology. Again, using our anthropic
probability, without assuming typicality or counting the number
of observers, this problem does not exist either.

If one is supposed to calculate the probability of a given theory based on
the data set D we collect, using Bayes' theorem:
\begin{equation}
P(T_i|D)={P(D|T_i)P(T_i)\over \sum_i P(D|T_i)P(T_i)},
\end{equation}
where $P(T_i)$ is a priori probability of theory $T_i$, then in
computing $P(D|T_i)$, our proposal eq.(3) will play a part. For a
given theory, the number of galaxies and the probability $p$ for
intelligence beings like us to develop are supposed to be
calculable. We need to try to avoid assuming typicality of us humans
in such a calculation, namely there is no reason to assume that we
are typical among all possible intelligent beings, thus $p$ must be
the probability for human beings to occur, not the probability for
just any kind of intelligence. As we shall discuss in the next
section, for most of applications, $p(p,n)$ is either 1 or 0, thus
according to Bayes' formula, as long as the data set boils down to
the minimum: only the existence of humans, then the posterior
probability of a given theory is either 0 or largely determined by
its a priori probability $P(T_i)$. Thus, for those theories which
can accommodate a large number of galaxies, anthropic reasoning
alone does not help to discriminate among them, only $P(T_i)$ figure
in the probability distribution, implying that theoretical
selectional principle is cried for if we are to be able to select
one or a few theories.

\section{Applications and examples}

\begin{figure}
\begin{center}
\includegraphics[width=4in]{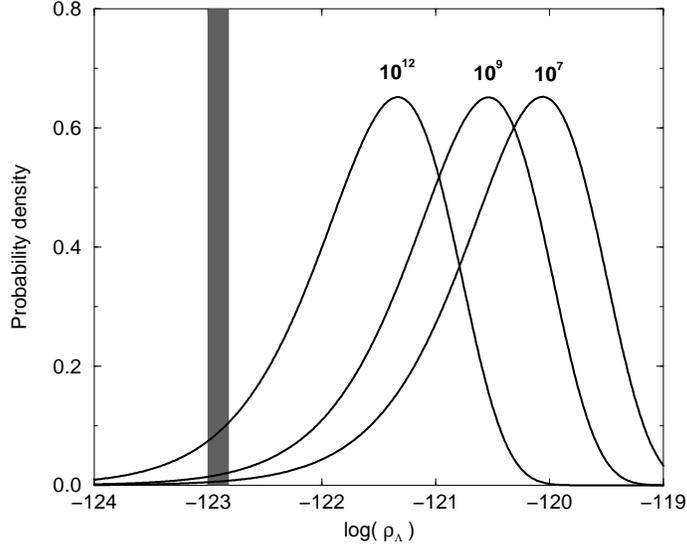}
\end{center}
\caption{\small Anthropic selection of the cosmological constant
using the baryon number weighted measure \cite{Bousso:2007kq}. Three curves
correspond to difference choices for the minimum
required mass for a galaxy: $M_* = (10^7, 10^9, 10^{12}) M_\odot$,
respectively.
The vertical bar corresponds to the observed value of the
cosmological constant. We find from this figure that the
cosmological constant we observe is not at the peak of the
probability distribution.} \label{peak}
\end{figure}

In addition to explaining paradoxes mentioned above, our anthropic
probability can be used to solve the problem that the observed
cosmological constant is not at the peak of the anthropic
probability distribution. As is shown in Fig. \ref{peak}, the
cosmological constant we observe is not the most probable value if
we use the baryon number weighted probability. However this problem
does not exist with our probability, it is because if the probability $p$ for
an observer to exist in one galaxy is not too small, and the number  of galaxies
$n$ is large enough, according to eq.(3) the anthropic probability $1-(1-p)^n$
can be very close to 1. So the distribution is more flat
and we find ourselves live in our meta-stable vacuum as it is  with a large anthropic
probability. Note that although the probability
distribution is sometimes more flat in our probability than in the
number weighted one, it is still good enough to explain the coincidences required
for observers' existence, because the anthropic probability
distribution usually falls exponentially when the deviation is
large from the most probable value, while our patch of the universe
can not be infinitely large \cite{eternalBH}.   Other solutions
to  this peak problem can be found in \cite{Bousso:2007kq}.

As another example of the  utility of our proposal, consider the
combined probability of the a priori probability and the anthropic
one. We calculate the probability in the ``ABZ'' model proposed by
Bousso \cite{Bousso:2006ev}. The reason for us to use this model is
that it is simple to calculate in this model, and the probability is
not number weighted. Also, this application can easily be
generalized to other models, where other a priori probabilities are
assigned.

\begin{figure}
\begin{center}
\includegraphics[width=4in]{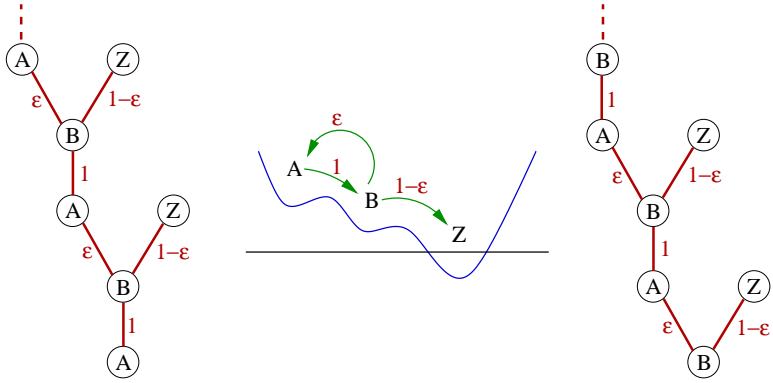}
\end{center}
\caption{\small A toy model of the landscape \cite{Bousso:2006ev}.
Assume that there are two de Sitter vacua $A$, $B$ and one terminal
AdS vacuum $Z$ in the landscape. $A$ can only tunnel to $B$ with
unit probability, $B$ can tunnel to $A$ with probability $\epsilon$,
or tunnel to $Z$ with probability $1-\epsilon$, and $Z$ can not
tunnel back to $A$ or $B$.} \label{abz}
\end{figure}

Assume there are only three vacua in the landscape, namely, the de
Sitter vacua $A$, $B$, and the AdS terminal vacuum $Z$. The decay
probabilities are shown in Fig. \ref{abz}. Suppose we start from the
vacuum $A$, i.e. the left figure in Fig. \ref{abz}, then the a
priori probability for the vacua are \cite{Bousso:2006ev}~

\begin{equation}
  q_A=\epsilon/2~, ~~ q_B=1/2~,~~ q_Z=(1-\epsilon)/(1+\epsilon)~,
\end{equation}
where we label the a priori probability by $q$ to to distinguish it
from the anthropic probability.

Note that the anthropic probability (\ref{ap}) jumps from zero to
one very fast if $n$ is exponentially large, where the exponential
is naturally produced by inflation. So we can approximate
$P_X(p,n)~(X=A, B, Z)$ by
\begin{equation}
  P_X(p,n)=
  \begin{cases}
    0& \text{when } np_X\ll 1, \text{ i.e. not suitable for observers},\\
    1& \text{when } np_X\gg 1, \text{ i.e. suitable for observers}.
  \end{cases}\label{bp}
\end{equation}

The full probabilities for vacua $A$, $B$ and $Z$ are proportional
to  $q_A P_A,~ q_B P_B, ~q_Z P_Z $ respectively. We see explicitly
in this example that the probability defined in this way is
well-behaved, finite, and does not suffer  the problems listed above
\footnote{At the first sight, it appears that problems such as the
freak observer problem do not arise here just because the ABZ model
itself does not suffer these problems. But it should be noticed that
the ABZ model itself only assigns a priori probability, and it must
be multipled by the anthropic probability to get the full
probability. If we do not use the anthropic probability properly,
the problems are still present}.

\section{Conclusion}

In this paper, we have argued that the exist-or-not anthropic
probability is a natural choice for eternal
inflation. This probability is free of the freak observer problem
and the typicality problem, it offers a better anthropic
explanation of the cosmological constant problem. A simple example
is discussed to demonstrate the utility of this probability.

The anthropic probability we propose can also be used in other
realizations for the landscape, or other kinds of the multiverses,
for a discussion of these realizations and multiverses, see
\cite{Weinberg:2005fh, Tegmark}.

It should be emphasized that although the anthropic probability
proposed in this paper is different from the number weighted
probability, we do not claim one of them is correct and
others are wrong. Different proposals come by due to different understanding of
the anthropic principle. As correctness of the anthropic
principle itself is not believed by everyone, it is still too early to
judge which version of the anthropic principle is correct. We only
suggest that our anthropic probability is natural, and do not
suffer the problems listed in this paper.

\section*{Acknowledgments}

This work was supported by grants of NSFC. We thank Yi-Fu Cai,
Chao-Jun Feng, Xian Gao, Wei Song, Yushu Song, Tower Wang, Zhiguang
Xiao, Wei Xue, and Xin Zhang for discussions. We also acknowledge R.
Bousso for allowing us to use the figures in his papers.

\end{document}